\documentclass[prd,twocolumn,superscriptaddress,amsmath,amssymb,nofootinbib]{revtex4-1}

\usepackage
{color}

\begin{document}

\title{First law and anisotropic Cardy formula for
three-dimensional Lifshitz black holes}

\author{Eloy Ay\'on-Beato}
\email{ayon-beato-at-fis.cinvestav.mx}
\affiliation{Departamento de F\'{\i}sica, CINVESTAV--IPN, Apdo.
Postal 14--740, 07000, M\'exico~D.F., M\'exico}
\affiliation{Instituto de Ciencias F\'isicas y Matem\'aticas,
Universidad Austral de Chile, Valdivia, Chile}

\author{Mois\'es Bravo-Gaete}
\email{mbravog-at-inst-mat.utalca.cl} \affiliation{Instituto de
Matem\'atica y F\'isica, Universidad de Talca, Casilla 747,
Talca, Chile}

\author{Francisco Correa}
\email{correa-at-cecs.cl} \affiliation{Leibniz Universit\"at
Hannover, Appelstra\ss{}e 2, 30167 Hannover, Germany}
\affiliation{Centro de Estudios Cient\'ificos (CECs), Valdivia,
Chile}

\author{Mokhtar~Hassa\"ine}
\email{hassaine-at-inst-mat.utalca.cl} \affiliation{Instituto
de Matem\'atica y F\'isica, Universidad de Talca, Casilla 747,
Talca, Chile}

\author{Mar\'ia Montserrat Ju\'arez-Aubry}
\email{mjuarez-at-fis.cinvestav.mx} \affiliation{Departamento
de F\'{\i}sica, CINVESTAV--IPN, Apdo. Postal 14--740, 07000,
M\'exico~D.F., M\'exico} \affiliation{Instituto de Ciencias
F\'isicas y Matem\'aticas, Universidad Austral de Chile,
Valdivia, Chile}
\affiliation{Instituto Tecnol\'ogico y de Estudios Superiores de Monterrey, Campus Puebla, V\'ia Atlixc\'ayotl No. 2301, Reserva Territorial Atlixc\'ayotl, Puebla. C.P. 72453 Puebla, M\'exico.}

\author{Julio Oliva}
\email{julio.oliva-at-uach.cl} \affiliation{Instituto de
Ciencias F\'isicas y Matem\'aticas, Universidad Austral de
Chile, Valdivia, Chile}

\begin{abstract}
The aim of this letter is to confirm in new concrete examples
that the semiclassical entropy of a three-dimensional Lifshitz
black hole can be recovered through an anisotropic
generalization of the Cardy formula, derived from the growth of
the number of states of a boundary non-relativistic field
theory. The role of the ground state in the bulk is played by
the corresponding Lifshitz soliton obtained by a double Wick
rotation. In order to achieve this task, we consider a scalar
field nonminimally coupled to new massive gravity for which we
study different classes of Lifshitz black holes as well as
their respective solitons, including new solutions for a
dynamical exponent $z=3$. The masses of the black holes and
solitons are computed using the quasilocal formulation of
conserved charges recently proposed by Gim \emph{et al.}
and based on the off-shell extension of the ADT formalism.  We
confirm the anisotropic Cardy formula for each of these
examples, providing a stronger base for its general validity.
Consistently, the first law of thermodynamics together with a
Smarr formula are also verified.
\end{abstract}

\maketitle

\section{Introduction}

During the last decade, there has been an intense activity to
promote the ideas underlying the gauge/gravity duality
\cite{Maldacena:1997re} in order to have a better understanding
of strongly coupled field theories with anisotropic scaling.
These latter are characterized by a scaling symmetry where
space and  time scale with different weights; their gravity
dual metric called the Lifshitz spacetime \cite{Kachru:2008yh}
is given by
\begin{equation}
ds^{2}=-\frac{r^{2z}}{l^{2z}}dt^2+\frac{l^2}{r^2}dr^2
+\frac{r^2}{l^2}d\vec{x}^2.\label{Lifshitz}
\end{equation}
Here $z$ is the dynamical critical exponent which reflects the
anisotropy of the scaling symmetry
\[
t\mapsto\tilde{\lambda}^z\,t,\qquad
r\mapsto\tilde{\lambda}^{-1}r,\qquad
\vec{x}\mapsto\tilde{\lambda}\,\vec{x}.
\]
Finite temperature effects are intended to be holographically
introduced through black holes commonly known as Lifshitz black
holes whose asymptotic behaviors match with the spacetime
(\ref{Lifshitz}). As it is now well-known, for $z\not=1$, in
order for the Einstein gravity to accommodate the Lifshitz
spacetimes some extra matter source is required like $p-$form
gauge fields, Proca fields
\cite{Kachru:2008yh,Taylor:2008tg,Pang:2009pd}, Brans-Dicke
scalars \cite{Maeda:2011jj} or eventually nonlinear
electrodynamic theories \cite{Alvarez:2014pra}. There also
exists the option of considering higher-order gravity theories
for which there are examples of Lifshitz black holes without
source, see
e.g.~\cite{AyonBeato:2009nh,Cai:2009ac,AyonBeato:2010tm,%
Matulich:2011ct,Oliva:2012zs,Giacomini:2012hg}. In this letter,
we will focus on a combination of these two options in three
dimensions by considering a gravity action given by a special
combination of quadratic curvature corrections to Einstein
gravity known as New Massive Gravity (NMG)
\cite{Bergshoeff:2009hq}, together with a source described by a
self-interacting scalar field nonminimally coupled to gravity.
We are then interested in the following three-dimensional
action {\small
\begin{align}
S[g,\Phi]={}&\frac{1}{2\kappa}\int\!\!d^{3}x\sqrt{-g}\Biggl[ R-2\lambda
-\frac{1}{m^{2}} \left( R_{\mu \nu }R^{\mu
\nu}-\frac{3}{8}R^{2}\right)  \Biggr]\nonumber\\
&-\int\!\!d^{3}x\sqrt{-g}\Biggl[
\frac{1}{2}\nabla_{\mu}\Phi\nabla^{\mu}\Phi+\frac{\xi}{2}R\Phi^2+U(\Phi)
\Biggr],
\label{action}
\end{align}}%
where $R$ denotes the scalar curvature, $R_{\mu\nu}$ the Ricci
tensor, $\xi$ stands for the nonminimal coupling parameter and
$U(\Phi)$ represents the self-interaction potential. There is a
variety of reasons that make this model worth of being
explored. Among others, it is well-known that nonminimally
coupled scalar fields are excellent laboratories in order to
evade standard ``no-hair'' theorems. In fact, it is known that
NMG supports a vacuum Lifshitz black hole for critical exponent
$z=3$ \cite{AyonBeato:2009nh} that is additionally
characterized as the only static axisymmetric asymptotically
Lifshitz solution that can be written as a Kerr-Schild
transformation of the Lifshitz spacetime (\ref{Lifshitz})
\cite{Ayon-Beato:2014wla}. Remarkably, it has been recently
shown that the spectrum of dynamical critical exponents $z$
supported by New Massive Gravity at finite temperature becomes
largely enriched if one includes self-interacting nonminimally
coupled scalar fields as sources of the Lifshitz black holes
according to the previous action \cite{Correa:2014ika}.
Besides, it is known that in standard three-dimensional Anti-de Sitter (AdS)
gravity supported by scalar fields, solitons play a fundamental
role as they can be treated as ground states; the existence of
scalar-tensor solitons turns out to be essential for
microscopically counting for the black holes entropy using
Cardy formula \cite{Correa:2010hf}. Here, we check that these
ideas also hold in the case of asymptotically Lifshitz
spacetimes in the presence of scalar fields, by carrying out
the microscopical computation of the black hole entropy using
the anisotropic generalization of the Cardy formula introduced
by Gonzalez \emph{et al.} \cite{Gonzalez:2011nz}.

As previously emphasized, the action (\ref{action}) allows
different classes of Lifshitz black hole configurations with
different values of the parameters and dynamical exponent. With
these solutions at hand, it is tempting to compute their
respective masses. However, the task of identifying the
conserved charges is a highly nontrivial problem whose
difficulty is increased in our case because of two main
reasons. Indeed, we are dealing with Lifshitz black holes which
have a rather nonstandard asymptotic behavior (\ref{Lifshitz})
and even more the gravity theory we consider (\ref{action})
contains quadratic corrections. In order to tackle this
problem, we will test the quasilocal formulation of conserved
charges recently proposed in \cite{Kim:2013zha,Gim:2014nba} and
based on the off-shell extension of the Abbott-Deser-Tekin (ADT) formalism
\cite{Abbott:1981ff}. In the ADT formalism, which is a covariant
generalization of the ADM method \cite{Arnowitt:1962hi}, the
metric $g_{\mu\nu}$ is linearized around the zero mass
spacetime with metric $\bar{g}_{\mu\nu}$ as
$g_{\mu\nu}=\bar{g}_{\mu\nu}+h_{\mu\nu}$. However, in this
approach, the perturbed metric $h_{\mu\nu}$ in the case of
Lifshitz black holes may not satisfy the correct assumptions
concerning the falloff boundary conditions, and consequently
yields an expression for the mass that does not satisfy the
first law of thermodynamics as shown in
\cite{Devecioglu:2010sf}. This problem has been recently
circumvented by the authors of \cite{Kim:2013zha,Gim:2014nba}
who have proposed a quasilocal generalization of the ADT
formalism that can be used even with slow falloff conditions.
In order to be as self-contained as possible and even if our
work is focused in the three-dimensional case, we briefly
present the main ingredients of the quasilocal ADT method in
arbitrary dimension $D$ following the notations and results of
\cite{Kim:2013zha,Gim:2014nba}. One of the interesting aspects
of this formalism lies in the fact that the computations are
done only on the basis of the full Lagrangian defining the
theory
$$
S[g,\Phi]={\int}d^Dx\sqrt{-g}\mathcal{L},
$$
and the Killing vectors $\xi^{\mu}$ associated to the
conserved charges without explicitly using the linearization of
the field equations. Here $\Phi$ collectively denotes any
matter content. The main result of Ref.~\cite{Kim:2013zha} is
the following prescription for the off-shell ADT potential
\begin{equation}
\sqrt{-g}{\cal Q}_{\mathrm{ADT}}^{\mu\nu}=
\frac{1}{2}\delta K^{\mu\nu}-\xi^{[\mu}\Theta^{\nu]},
\end{equation}
in terms of the surface term $\Theta^{\mu}$ arising from the
variation of the action
$$
\delta{S}={\int}d^Dx[\sqrt{-g}(\mathcal{E}_{\mu\nu}{\delta}g^{\mu\nu}
+\delta_\Phi\mathcal{L})+\partial_\mu\Theta^{\mu}],
$$
and the off-shell Noether potential $K^{\mu\nu}$ associated to
the identical conservation of the off-shell Noether current
$$
J^\mu=\sqrt{-g}(\mathcal{L}g^{\mu\nu}+2\mathcal{E}^{\mu\nu})\xi_\nu
-\Theta^{\mu}=\partial_{\nu}K^{\mu\nu}.
$$
It is important to remark that the exclusive use of off-shell
conserved currents in the derivation makes that the background
metric is not required to satisfy the field equations. Other
important fact is that the standard linearization method is
only compatible at the asymptotic regime. Hence, in order to
circumvent this problem and to construct quasilocal charges,
the authors in \cite{Kim:2013zha,Gim:2014nba} consider
linearizing along a one-parameter family of configurations as
has been advocate, for example, in \cite{Wald:1999wa}. More
concretely, for the configurations under study a parameter
$0\le{s}\le1$ can be introduced which allows the interpolation
with the asymptotic solution at $s=0$. In doing so, the
quasilocal conserved charge reads
\begin{equation}\label{eq:5}
Q(\xi)=\int_{\cal B}\!d^{D-2}x_{\mu\nu}
\Big(\Delta{K}^{\mu\nu}(\xi)-2\xi^{[\mu}\!\!
\int^1_0\!\!ds~\Theta^{\nu]}(\xi|s)\Big),
\end{equation}
where $\Delta{K}^{\mu\nu}(\xi)\equiv
{K}^{\mu\nu}_{s=1}(\xi)-K^{\mu\nu}_{s=0}(\xi)$ denotes the
difference of the Noether potential between the interpolated
solutions, and $d^{D-2}x_{\mu\nu}$ represents the integration
over the co-dimension two boundary ${\cal B}$. In our case
(\ref{action}), the involved quantities are given by
\begin{align}
\Theta^\mu={}&2\sqrt{-g}\biggl(P^{\mu\alpha
\beta\gamma}\nabla_\gamma\delta g_{\alpha\beta}
-\delta g_{\alpha\beta}\nabla_\gamma P^{\mu\alpha\beta\gamma}\nonumber\\
&\qquad\quad
+\frac{1}{2}
\frac{\partial\mathcal{L}}{\partial\left(\partial_{\mu}\Phi\right)}
\delta\Phi\biggr), \label{eq:theta}\\
K^{\mu\nu}={}&\sqrt{-g}\left(2P^{\mu\nu\rho\sigma}\nabla_\rho\xi_\sigma
-4\xi_\sigma\nabla_\rho P^{\mu\nu\rho\sigma}\right) \label{eq:K},
\end{align}
with $P^{\mu\nu\sigma\rho}\equiv{\partial\mathcal{L}}/{\partial
{R}_{\mu\nu\sigma\rho}}$. Notice that we have added the
contribution to the surface term coming from the scalar field,
which was not included in the original
Refs.~\cite{Kim:2013zha,Gim:2014nba} since they dealt with the
vacuum case.
It is worth to note that generic matter contributions to the quasilocal ADT formalism has been investigated in \cite{Hyun:2014nma}.

After computing the masses of our solutions, we subject the
quasilocal ADT method to two nontrivial tests. First, we will
verify that the first law of thermodynamics holds in each case
by computing the Wald formula \cite{Wald:1993nt} for the
entropy as
\begin{equation}
\mathcal{S}_{\mathrm{W}}=-2\pi\,\Omega_{D-2}\,
\left(\frac{r_{h}}{l}\right)^{D-2}\,
\left[P^{\alpha\beta\mu\nu}\,\varepsilon_{\alpha\beta}\,
\varepsilon_{\mu\nu}\right]_{r=r_{h}},
\label{w2}
\end{equation}
together with the Hawking temperature $T$; here $r_h$ stands
for the location of the horizon. Independently, in three
dimensions a Lifshitz black hole characterized by a dynamical
exponent $z$ and mass $\mathcal{M}$ has a corresponding soliton
with dynamical exponent $z^{-1}$ and mass
$\mathcal{M}_{\mathrm{sol}}$ obtained by operating a double
Wick rotation \cite{Gonzalez:2011nz}. We will confirm that our
mass expressions are compatible with the anisotropic
generalization of the Cardy formula proposed in
\cite{Gonzalez:2011nz}
\begin{equation}\label{generCardy}
\mathcal{S}_{\mathrm{C}}=2\pi{l}(z+1)\left[
\left(-\frac{\mathcal{M}_{\mathrm{sol}}}{z}\right)^{z}\mathcal{M}\right]
^{\frac{1}{z+1}}.
\end{equation}
This formula is obtained starting from a boundary
non-relativistic field theory and computing the asymptotic
growth of the number of states with fixed energy, assuming the
role of ground state in the bulk is to be played by the
soliton. Notice that for AdS asymptotics, $z=1$, this formula
becomes exactly the Cardy formula used in \cite{Correa:2010hf}
for the case of standard gravity supported by scalar fields.
Remarkably, both tests will be checked satisfactorily by the
quasilocal ADT method.

The rest of the paper is organized as follows. In the next
section, we will study different classes of Lifshitz black hole
solutions for which we will obtain their mass using the
quasilocal ADT formalism. We explicitly verify the fulfilment
of the first law in all these cases.  Moreover, for each
Lifshitz black hole solution, we obtain their corresponding
soliton counterpart, compute their mass and spotlight the
validity of the anisotropic Cardy formula (\ref{generCardy}).
Finally, the last section is devoted to our conclusions where
we also anticipate the generalization of some of these results
to higher dimensions.

\section{Nonminimally dressed Lifshitz black holes}

The field equations obtained by varying the action
(\ref{action}) read
\begin{subequations}\label{eq:NMGS}
\begin{align}
G_{\mu\nu}+\lambda{g}_{\mu\nu}-\frac{1}{2m^2}K_{\mu\nu}={}&{\kappa}T_{\mu\nu},
\label{eqmotion1}\\
\Box\Phi-{\xi}R\Phi ={}& \frac{dU(\Phi)}{d\Phi},\label{eqmotion2}
\end{align}
where the higher-order contribution $K_{\mu\nu}$ and the
nonminimally coupled energy-momentum tensor $T_{\mu\nu}$ are
defined by
\begin{align}
K_{\mu\nu}={}&2\Box R_{\mu\nu} -\frac{1}{2}
\left( g_{\mu\nu}\Box+\nabla_{\mu}\nabla_{\nu}-9R_{\mu\nu} \right)R
\nonumber\\
&-8R_{\mu\alpha}R^{\alpha}_{~\nu}
+g_{\mu\nu}\left( 3R^{\alpha\beta}R_{\alpha\beta}-\frac{13}{8}R^2 \right),\\
T_{\mu\nu}={}&\nabla_{\mu}\Phi\nabla_{\nu}\Phi
-g_{\mu\nu}\left(\frac{1}{2}\nabla_{\sigma}\Phi\nabla^{\sigma}\Phi+U(\Phi)\right)\nonumber\\
&+\xi( g_{\mu\nu}\Box - \nabla_{\mu}\nabla_{\nu}+G_{\mu\nu} )\Phi^2.
\end{align}
\end{subequations}

In what follows, we will consider Lifshitz black holes within
the following ansatz
\begin{equation}
ds^2=-\frac{r^{2 z}}{l^{2z}} f(r) dt^2 + \frac{l^2}{r^2}\,\frac{dr^2}{f(r)}
+ \frac{r^2}{l^2}d{\varphi}^2,\label{lifbh}
\end{equation}
with coordinates ranges defined by $-\infty<t<\infty$,
$0<r<\infty$ and the angular variable $0\leq\varphi<2\pi{l}$.
We start by reanalyzing one of the solutions obtained by some
of the authors in Ref.~\cite{Correa:2014ika}, the single one
characterized by having a non-vanishing Wald entropy, and
therefore suitable for our analysis. For the exponent $z=3$ the
configurations of Ref.~\cite{Correa:2014ika} only contain the
vacuum black hole of NMG \cite{AyonBeato:2009nh}, for this
reason we later concentrate on this value of the exponent and
exhibit nonminimally dressed Lifshitz black holes also in this
case.

\subsection{Nonminimally dressed black holes for generic $z$}

The following family of Lifshitz black hole solutions was found
in Ref.~\cite{Correa:2014ika}
\begin{subequations}\label{lifbhsoln1}
\begin{align}
ds^2  ={}&-\frac{r^{2z}}{l^{2z}}
           \left[1-M\left(\frac{l}{r}\right)^{\frac{z+1}{2}}\right]dt^2
           \nonumber\\
         &+\frac{l^2}{r^2}
           \left[1-M\left(\frac{l}{r}\right)^{\frac{z+1}{2}}\right]^{-1}{dr^2}
          +\frac{r^2}{l^2}d{\varphi}^2,\\
\Phi(r)={}&\sqrt{\frac{(z-3)(9z^2-12z+11)M}{2\kappa(z-1)(z^2-3z+1)}}
           \left(\frac{l}{r}\right)^{\frac{z+1}{4}},
\end{align}
where the scalar field has the self-interaction
\begin{align}\label{pot1}
U(\Phi)={}&\frac{(z-1)(21z^3-13z^2+31z-15)}{32l^2(9z^2-12z+11)}\Phi^2
\nonumber\\
&-\frac{(z-1)^3(z^2-3z+1)(9z^2-12z+19)\kappa}
{32l^2(z-3)(9z^2-12z+11)^2}\Phi^4,
\end{align}
and the coupling constants are parameterized by
\begin{align}
m^2&=-\frac{z^2-3z+1}{2l^2},&
\lambda=-\frac{z^2+z+1}{2l^2}, \nonumber\\
\xi&=\frac{3z^2-4z+3}{2(9z^2-12z+11)}.
\label{spp}
\end{align}
\end{subequations}
Before proceeding with the computations of the Noether
potential and the surface term in order to derive the mass, we
would like to emphasize some aspects of this solution. First of
all, because of the expression of the scalar field, this
solution has no AdS limit $z=1$. The other special value of the
dynamical exponent is given by $z=3$, for which the scalar
field as well as the potential vanish identically, and one ends
with the vacuum black hole of NMG \cite{AyonBeato:2009nh}. We
notice from now that for this class of solution, as well as for
the two other solutions derived below, the allowed potential
always involves a mass term.

It is easy to see that this class of solution has a
nonvanishing Wald entropy and a Hawking temperature given by
\begin{align}\label{entropytemsoln1}
\mathcal{S}_{\mathrm{W}}&=
-\frac{\pi^2(z+1)^2(3z-5)r_{h}}{2\kappa(z-1)(z^2-3z+1)},\\
T&=\frac{(z+1){r_h}^z}{8{\pi}l^{z+1}}, & r_h=l\,M^{\frac{2}{z+1}}.
\end{align}
Let us now compute the mass through the quasilocal ADT
formalism. For the timelike Killing vector $\xi^t=(1,0,0)$, and
after some tedious but straightforward computations, the
expressions for the Noether potential and the surface term are
given by {\small
\begin{align}
\int_{0}^{M}\!\!\!\!dM\Theta^{r}={}&
\frac{(z+1)(9z^3-31z^2+31z-25)M}{16{\kappa}l(z-1)(z^2-3z+1)}
\left(\frac{r}{l}\right)^{\frac{z+1}{2}}\nonumber\\
&-\frac{(z+1)(15z^3-60z^2+67z-50)M^2}{32{\kappa}l(z-1)(z^2-3z+1)},
\nonumber\\
K^{rt}={}&-\frac{(z+1)(9z^3-31z^2+31z-25)M}{16{\kappa}l(z-1)(z^2-3z+1)}
\left(\frac{r}{l}\right)^{\frac{z+1}{2}}\nonumber\\
&\frac{3(z+1)(z-3)(5z^2-6z+5)M^2}{32{\kappa}l(z-1)(z^2-3z+1)}.
\nonumber
\end{align}}%
This implies that the mass $\mathcal{M}$ of the Lifshitz black
hole solution (\ref{lifbhsoln1}) turns to be
\begin{equation}\label{masssoln1}
\mathcal{M}=-\frac{\pi(z+1)^2(3z-5)}{16\kappa(z-1)(z^2-3z+1)}
\left(\frac{r_h}{l}\right)^{z+1}.
\end{equation}
It is simple to verify that the black hole entropy
(\ref{entropytemsoln1}) and the mass (\ref{masssoln1}) satisfy
the first law of black hole thermodynamics
$d\mathcal{M}=T\,d\mathcal{S}_{\mathrm{W}}$. In fact, they
satisfy an anisotropic version of the Smarr formula
\begin{equation}\label{Smarr}
\mathcal{M}=\frac{T}{z+1}\,\mathcal{S}_{\mathrm{W}},
\end{equation}%
\begin{table}
\caption{\label{tabla1}Range of possibilities for the dynamical
exponent $z$ allowing positive mass black holes,
$\mathcal{M}>0$.}
\begin{tabular}{|c|c|}
\hline
$\kappa$ & Range of $z$ \\
\hline \hline
~$\kappa>0$~ & ~$1.7\approx5/3<z<{(3+\sqrt{5})}/{2}\approx2.6$~ \\  [1ex]
\hline \hline
~$\kappa<0$~ & ~$2.6\approx(3+\sqrt{5})/2<z\leq3$~ \\  [1ex]
\hline
\end{tabular}
\end{table}%
which stipulates that the mass $\mathcal{M}$ as function of the
entropy $\mathcal{S}_{\mathrm{W}}$ is a homogeneous function of
degree $z+1$ \cite{Smarr:1972kt}. Notice that these nice
properties are satisfied independently of the sign of the mass.

It is also interesting to note that for the vacuum case $z=3$
\cite{AyonBeato:2009nh}, this expression of the mass coincides
with the one derived in different papers using others
formalisms
\cite{Cai:2009ac,Myung:2009up,Hohm:2010jc,Gonzalez:2011nz},
provided that the Einstein constant is taken negative
$\kappa=-8{\pi}G$; i.e.\ by choosing the so-called ``wrong''
sign of NMG which turns tensor ghosts at linearized level into
unitary Fierz-Pauli massive excitations on maximally symmetric
vacua \cite{Bergshoeff:2009hq}. Clearly, the range of the
dynamical exponent $z$ which ensures a positive mass strongly
depends on the sign of the coupling constant $\kappa$. Imposing
the positivity of the mass and the reality of the scalar field,
the critical exponent $z$ must be restricted according to Table
\ref{tabla1}.

We now derive the corresponding soliton solution which exists
for the same range of parameters and self-interacting potential
than the black hole solution (\ref{lifbhsoln1}). This soliton
solution turns out to have a dynamical exponent $z^{-1}$ and a
characteristic scale $lz^{-1}$ which is a consequence of the
two-dimensional isomorphism between the Lifshitz Lie algebras
with dynamical exponents $z$ and $z^{-1}$ obtained by swaping
the role of the Hamiltonian with the momentum generator
\cite{Gonzalez:2011nz}. We will present in details the
different steps in this case and only report the main results
in the other two solutions. We first consider the Euclidean
version of the Lifshitz black hole (\ref{lifbhsoln1}) obtained
by the Wick rotation $t=i\tau$,
\begin{equation}\label{lifbhsoln2}
ds^2 = \frac{r^{2z}}{l^{2z}}f(r)d\tau^2
 + \frac{l^2}{r^2f(r)}{dr^2} + \frac{r^2}{l^2}d{\varphi}^2,
\end{equation}
where the metric function $f(r)$ can be read from
Eq.~(\ref{lifbhsoln1}) and the static scalar field remains the
same. In order to avoid conical singularities, the Euclidean
time must be periodic with period $\beta=T^{-1}$, that is
$0\leq\tau<\beta$ and the angle keeps identified as
$0\leq\varphi<2{\pi}l$. Under the Euclidean diffeomorphism
defined by
\begin{equation}\label{diffeo}
(\tau,r,\varphi)\mapsto\left(\bar{\tau}=\left(\frac{2\pi l}{\beta}\right)^{\frac{1}{z}}\varphi,
\bar{r}=\frac{\beta}{2\pi z}\left(\frac{r}{l}\right)^z,
\bar{\varphi}=\frac{2\pi l}{\beta}\tau\right),
\end{equation}
the line element (\ref{lifbhsoln2}) becomes
\begin{align}
ds^2&=\left(\frac{z\bar{r}}{l}\right)^{\frac{2}{z}}d\bar{\tau}^2
+\frac{l^2}{z^2\bar{r}^2 F(\bar{r})}d\bar{r}^2
+\frac{z^2 \bar{r}^2}{l^2}F(\bar{r})d\bar{\varphi}^2,\nonumber\\
F(\bar{r})&=1-\bar{M}\left(\frac{l}{z\bar{r}}\right)^{\frac{z+1}{2z}},\quad
\bar{M}=\left(\frac{4}{z+1}\right)^{\frac{z+1}{2z}},
\end{align}
i.e.\ the Euclidean Lifshitz black hole is diffeomorphic to
another asymptotically Lifshitz solution with dynamical
exponent $z^{-1}$, scale $lz^{-1}$ and temperature
\begin{equation}
\bar{\beta}=\left(2\pi l\right)^{1+\frac{1}{z}}\beta^{-\frac{1}{z}}.
\end{equation}
It is, in fact, a soliton, its regular character is not manifest
in these coordinates which have the advantage of exposing the
Lifshitz asymptotic behavior with exponent $z^{-1}$ and scale
$lz^{-1}$. Finally, the corresponding Lorentzian soliton
obtained through $\bar{\tau}=i\bar{t}$ reads
\begin{subequations}\label{lifsolsoln1}
\begin{align}
ds^2&=-\left(\frac{z\bar{r}}{l}\right)^{\frac{2}{z}}d\bar{t}^2
+\frac{l^2}{z^2\bar{r}^2 F(\bar{r})}d\bar{r}^2
+\frac{z^2\bar{r}^2}{l^2}F(\bar{r})d\bar{\varphi}^2,\\
\Phi(\bar{r})&=\sqrt{\frac{(z-3)(9z^2-12z+11)\bar{M}}{2\kappa(z-1)(z^2-3z+1)}}
\left(\frac{l}{z\bar{r}}\right)^{\frac{z+1}{4z}}.
\end{align}
\end{subequations}
Now promoting the fixed constant to a variable one,
$\bar{M}{\mapsto}s\bar{M}$, we obtain a one-parameter family of
local solutions which facilitates the computation of the mass
along the same lines as before. Before proceeding with the
computation of the mass, let us analyze this last point
carefully. For the black hole solution, since the constant $M$
is an integration constant, a parameter $s$ with range $s\in
[0,1]$ could have been introduced in the solution via the
change $M{\mapsto}sM$. This change is useful only for computing
the mass since in this case, the variation will be operated
with the parameter $s$, and the surface term will be rather
integrated as $\int_0^1 ds\,\Theta^r$. Nevertheless, the result
is the same if one promotes the constant $M$ as the moving
parameter and integrating the surface term from $0$ to $M$ as
we did previously. Now, for the counterpart soliton, one can
start with the black hole solution parameterized in term of $s$
and operate the same diffeomorphism (\ref{diffeo}). The
resulting soliton solution will correspond to the solution
(\ref{lifsolsoln1}) with $\bar{M}{\mapsto}s\bar{M}$, and the
integration of the surface term will be given by $\int_0^1
ds\,\Theta^{\bar{r}}$. However, as in the black hole case, the
new variable constant $\bar{M}$ can be used as the moving
parameter, and the result is exactly the same. Hence, choosing
the Killing vector as $\xi^{\bar{t}}=(1,0,0)$, the Noether
potential and the surface term take the following form {\small
\begin{align}
\int_{0}^{\bar{M}}\!\!\!\!d\bar{M}\,\Theta^{\bar{r}}={}&
\frac{(z+1)(9z^3-31z^2+31z-25)\bar{M}}{16{\kappa}l(z-1)(z^2-3z+1)}
\left(\frac{z\bar{r}}{l}\right)^{\frac{z+1}{2z}}\nonumber\\
&-\frac{(z+1)(15z^3-60z^2+67z-50)\bar{M}^2}{32{\kappa}l(z-1)(z^2-3z+1)},
\nonumber
\end{align}
\begin{equation}
K^{\bar{r}\bar{t}}=\frac{(z+1)(9z^3-31z^2+31z-25)\bar{M}}
{16{\kappa}l(z-1)(z^2-3z+1)}
\left[\bar{M}-\left(\frac{z\bar{r}}{l}\right)^{\frac{z+1}{2z}}\right],
\nonumber
\end{equation}}%
giving a unique value for the mass of the Lifshitz soliton,
independent of any integration constant as expected,
\begin{eqnarray}\label{massolsol1}
\mathcal{M}_{\mathrm{sol}}=\frac{{\pi}z(3z-5)}{\kappa(z-1)(z^2-3z+1)}
\left(\frac{z+1}{4}\right)^{\frac{z-1}{z}}.
\end{eqnarray}
It is straightforward to check that the mass of the soliton
and the mass of the black hole have opposite signs
(\ref{masssoln1}) as expected. As long as the mass of the
soliton is negative $\mathcal{M}_{\mathrm{sol}}<0$ (see Table
\ref{tabla1}), the holographic picture unveiled in
Ref.~\cite{Gonzalez:2011nz} applies: the semiclassical entropy
of the Lifshitz black hole (\ref{lifbhsoln1}) can be understood
from the asymptotic growth of the number of states of a $1+1$
non-relativistic field theory with ground state corresponding
in the bulk to the soliton (\ref{lifsolsoln1}). This gives rise
to the anisotropic generalization of the Cardy formula
(\ref{generCardy}) which after evaluation perfectly coincides
with the Wald formula (\ref{entropytemsoln1})
\begin{equation}
\mathcal{S}_{\mathrm{W}}=\mathcal{S}_{\mathrm{C}}\, .
\end{equation}

We would like to emphasize that the bulk semiclassical
derivation of Ref.~\cite{Gonzalez:2011nz} is also applicable to
negative mass black holes (positive mass solitons), which as we
already show are compatible with the first law. In this case
the anisotropic formula involves the absolute values of the
masses and consequently with the first law produces a negative
entropy; i.e.\ a general formula would be
\begin{equation}\label{|Cardy|}
\mathcal{S}_{\mathrm{C}}=\epsilon\,2\pi{l}(z+1)\left[
\left(\frac{|\mathcal{M}_{\mathrm{sol}}|}{z}\right)^{z}|\mathcal{M}|\right]
^{\frac{1}{z+1}},
\end{equation}
where $\epsilon=\pm1$ corresponds to the sign of the black hole
mass. Obviously, for $\epsilon=-1$ an holographic
interpretation has no sense; even the mere existence of a
thermodynamical one can be challenged.

Moreover, for the vacuum dynamical exponent $z=3$ and with
$\kappa=-8 \pi G$, the soliton mass (\ref{massolsol1}) becomes
$\mathcal{M}_{\mathrm{sol}}=-3/(4G)$, which precisely
corresponds to the mass of the vacuum gravitational soliton
found in \cite{Gonzalez:2011nz}. In the following sections we
exhibit new nonminimally dressed Lifshitz solutions for the
same exponent $z=3$, and since the involved steps are similar
only the important results are reported.

\subsection{\label{family2_3d}Nonminimally dressed black holes for $z=3$}

For dynamical exponent $z=3$ a new family of Lifshitz black
holes solutions is presented
\begin{subequations}\label{eq:family}
\begin{align}
ds^2={}&-\frac{r^6}{l^6}\left(1-\frac{Ml^4}{r^4}\right)dt^2
      +\frac{l^2}{r^2}\left(1-\frac{Ml^4}{r^4}\right)^{-1}dr^2\nonumber\\
&+\frac{r^2}{l^2}d\varphi^2, \\
\Phi(r)={}&\sqrt{\frac{M}{\kappa(2-13\xi)}}\frac{l^2}{r^2},
\end{align}
this configuration is supported by the self-interacting
potential
\begin{equation}
U(\Phi)=-\frac{2-13\xi}{2l^2}\left[2\Phi^2+(2-\xi)\kappa\Phi^4\right],
\end{equation}
where the nonminimal coupling parameter $\xi$ is not restricted
\emph{a priori} and the remaining coupling constants are related as in
vacuum
\begin{equation}\label{eq:lambda-m^2z=3}
\frac{\lambda}{13}=m^2=-\frac{1}{2l^2}.
\end{equation}
\end{subequations}
As before, choosing the Killing vector $\xi^{t}=(1,0,0)$, the
Noether potential and the surface term are calculated as
\begin{align}
\int_{0}^{M}\!\!\!\!dM\Theta^{r}&=
-\frac{2(4\xi-1)M^2l^3}{\kappa(13\xi-2)r^4}
+\frac{(104\xi-17)M}{{\kappa}l(13\xi-2)},\nonumber\\
K^{rt}&=\frac{2(4\xi-1)M^2l^3}{\kappa(13\xi-2)r^4}
-\frac{2(68\xi-11)M}{{\kappa}l(13\xi-2)},\nonumber
\end{align}
from which we obtain the mass of the Lifshitz black hole
(\ref{eq:family}) as
\begin{equation}\label{masssoln2}
\mathcal{M}=-\frac{2\pi(32\xi-5)}{\kappa(13\xi-2)}
\left(\frac{r_h}{l}\right)^4.
\end{equation}
It is easy to see that this expression for the mass satisfies
the first law, since the related Wald entropy and temperature
are expressed by
\begin{align}\label{entropytemsoln2}
\mathcal{S}_{\mathrm{W}}&=-\frac{8\pi^2(32\xi-5)r_h}{\kappa(13\xi-2)},\\
T&=\frac{{r_h}^3}{{\pi}l^4}, & r_h=l M^{1/4},
\end{align}%
\begin{table}
\caption{\label{tabla2}Range of possibilities for the
nonminimal coupling parameter $\xi$.}
\begin{tabular}{|c|c|c|}
\hline
$\kappa$ & ~Range of $\xi$~ & ~$\mathcal{M}>0$~ \\
\hline \hline
~$\kappa>0$~ & ~$\xi<2/13\approx0.154$~ & $ \emptyset $ \\  [1ex]
\hline \hline
~$\kappa<0$~ & ~$\xi>2/13\approx0.154$~ & ~$\xi>5/32\approx0.156$~ \\  [1ex]
\hline
\end{tabular}
\end{table}%
which is also compatible with the Smarr formula (\ref{Smarr})
for $z=3$. The possible values for the nonminimal coupling
parameter warranting the existence of the solution and from
them those allowing positive mass are all summarized in Table
\ref{tabla2}.

As in the previous case, operating the same diffeomorphism
(\ref{diffeo}) with $z=3$ on the Euclidean version of the
solution (\ref{eq:family}), we obtain a Lifshitz soliton whose
Lorentzian counterpart is
\begin{subequations}\label{eq:family_sol}
\begin{align}
ds^2 ={}&
-\left(\frac{3\,\bar{r}}{l}\right)^{2/3} d\bar{t}^2
+\frac{l^2}{9\,\bar{r}^2}
\left[1-\bar{M}\left(\frac{l}{3\,\bar{r}}\right)^{4/3}\right]^{-1}d\bar{r}^2\nonumber\\
&+\frac{9\,\bar{r}^2}{l^2}
\left[1-\bar{M}\left(\frac{l}{3\,\bar{r}}\right)^{4/3}\right]d{\bar{\varphi}}^2,\\
\Phi(\bar{r})={}&\sqrt{\frac{\bar{M}}{\kappa(2-13\xi)}}
\left(\frac{l}{3\,\bar{r}}\right)^{2/3}, \quad \bar{M}=2^{-4/3}.
\end{align}
\end{subequations}
Once again, introducing a one-parameter family of locally
equivalent solutions via $\bar{M}{\mapsto}s\bar{M}$, the
Noether potential and the surface term are obtained as
\begin{align}
\int_{0}^{\bar{M}}\!\!\!\!d\bar{M}\Theta^{\bar{r}}&=
-\frac{2(4\xi-1)\bar{M}^2}{{\kappa}l(13\xi-2)}\!
\left(\frac{l}{3\,\bar{r}}\right)^{4/3}
+\frac{(104\xi-17)\bar{M}}{{\kappa}l(13\xi-2)},\nonumber\\
K^{\bar{r}\bar{t}}&=
\frac{2(4\xi-1)\bar{M}^2}{{\kappa}l(13\xi-2)}
\left(\frac{l}{3\,\bar{r}}\right)^{4/3}
-\frac{2(4\xi-1)\bar{M}}{{\kappa}l(13\xi-2)},\nonumber
\end{align}
which in turn implies the following fixed mass for the soliton
(\ref{eq:family_sol})
\begin{equation}\label{massolsol2}
\mathcal{M}_{\mathrm{sol}}=\frac{3\pi(32\xi-5)}{2^{1/3}\kappa(13\xi-2)}.
\end{equation}
For $\xi>5/32$ and $\kappa<0$, it is straightforward to check
that the generalized Cardy formula (\ref{generCardy}) fits
perfectly with the expressions of the masses of the Lifshitz
black hole (\ref{masssoln2}), its soliton counterpart
(\ref{massolsol2}) and the Wald entropy
(\ref{entropytemsoln2}).

Here again, the cases with negative black hole masses, that can
be inferred from Table \ref{tabla2} i.e.\ $\xi<2/13$ for
$\kappa>0$ and $2/13<\xi<5/32$ for $\kappa<0$, are compatible
with the first law, the Smarr formula (\ref{Smarr}), and its
entropy can be rewritten \emph{\`a la} Cardy according to the
general formula (\ref{|Cardy|}) without further interpretation.
In fact, for one of these nonminimal couplings, namely
$\xi=3/20<2/13$ with $\kappa>0$, the solution can be improved
by generalizing the self-interaction with the addition of a
cubic contribution. The result is a sort of rigid dressing of
the vacuum $z=3$ black hole \cite{AyonBeato:2009nh}.

\subsection{\label{family3_3d}Dressing the vacuum $z=3$ black hole for
$\xi=3/20$}

In the scenario where the nonminimal coupling takes the value
$\xi=3/20$ and $\kappa>0$, the solution (\ref{eq:family}) is
improved to
\begin{subequations}\label{fam3}
\begin{align}
ds^2={}&-\frac{r^6}{l^6}
\left(1-\frac{\alpha\sqrt{M}l^2}{r^2}-\frac{Ml^4}{r^4}\right)dt^2
\nonumber\\
&+\frac{l^2}{r^2}
\left(1-\frac{\alpha\sqrt{M}l^2}{r^2}-\frac{Ml^4}{r^4}\right)^{-1}dr^2
+ \frac{r^2}{l^{2}}d\varphi^2,
\label{metricfamily3} \\
\Phi(r)={}&\sqrt{\frac{20M}{\kappa}}\frac{l^2}{r^2},
\end{align}
provided that the self-interaction potential is generalized as
\begin{equation}
U(\Phi) = -\frac{1}{20l^2}\Phi^2 -
\frac{\alpha\sqrt{\kappa}}{5\sqrt{5}l^2}\Phi^3
-\frac{37\kappa}{800l^2}\Phi^4,
\end{equation}
\end{subequations}
with no restrictions in the cubic coupling constant $\alpha$
and the remaining coupling constants fixed as in
(\ref{eq:lambda-m^2z=3}). Additionally to the $\alpha=0$ limit,
where we consistently recover the black hole solution
(\ref{eq:family}) for $\xi=3/20$, this solution allows another
nontrivial limit: for $M\rightarrow0$ and
$\alpha\rightarrow\infty$ keeping fixed the quantity
$M_v=\alpha\sqrt{M}$ this solution becomes just the vacuum
$z=3$ black hole \cite{AyonBeato:2009nh} with integration
constant $M_v$. This solution can be interpreted as a sort of
rigid dressing of the vacuum $z=3$ black hole by a
self-interacting scalar field with nonminimal coupling
$\xi=3/20$.

Calculating the Wald entropy and temperature of this black hole
gives
\begin{align}
\mathcal{S}_{\mathrm{W}}&=-\frac{32\pi^2\sqrt{\alpha^2+4}\,r_h}
{\kappa\left(\alpha+\sqrt{\alpha^2+4}\right)},\label{entropysoln3}\\
T&=\frac{\sqrt{\alpha^2+4}\,{r_h}^3}
{{\pi}l^4\left(\alpha+\sqrt{\alpha^2+4}\right)},
& {r_h}^2 = \frac{l^2\sqrt{M}}2\left(\alpha+\sqrt{\alpha^2+4}\right).
\label{temsoln3}
\end{align}
For the same timelike Killing vector, the expressions of  the
Noether potential and the surface term read 
\begin{align*}
\int_{0}^{M}\!\!\!\!dM\Theta^{r}={}&\frac{4\alpha\sqrt{M}r^2}{{\kappa}l^3}
-\frac{(\alpha^2-28)M}{{\kappa}l}
-\frac{20l\alpha{M}^{3/2}}{\kappa{r}^{2}}\\
&-\frac{16{l}^{3}{M}^{2}}{\kappa{r}^{4}},\\
K^{rt}={}&\frac{16{l}^{3}{M}^{2}}{\kappa{r}^{4}}
+\frac{20l\alpha{M}^{3/2}}{\kappa{r}^{2}}
-\frac{32M}{{\kappa}l}
-\frac{4\alpha\sqrt{M}{r}^{2}}{\kappa{l}^{3}},
\end{align*}
giving the mass
\begin{equation}\label{masssoln3}
\mathcal{M}=-\frac{8\pi(\alpha^2+4)}
{\kappa\left(\alpha+\sqrt{\alpha^2+4}\right)^2}\left(\frac{r_h}{l}\right)^4.
\end{equation}
As in the previous examples, the first law is satisfied, as well
as the Smarr formula. Notice that the cubic interaction does
not enhance the sign of the mass, which remains negative as in
the  case with $\alpha=0$.

Following the same lines of the two previous examples, the
corresponding soliton reads
\begin{subequations}\label{soliton3}
\begin{align}
ds^2&=-\left(\frac{3\,\bar{r}}{l}\right)^{2/3} d\bar{t}^2
+\frac{l^2}{9\,\bar{r}^2}\frac{d\bar{r}^2}{F(\bar{r})}
+ \frac{9\,\bar{r}^2}{l^{2}}F(\bar{r})\,d\bar{\varphi}^2,\\
\Phi(\bar{r})&=\sqrt{\frac{20\bar{M}}{\kappa}}\left(\frac{l}{3\,\bar{r}}\right)^{2/3},\\
F(\bar{r})&=1-\alpha\,\sqrt{\bar{M}}\left(\frac{l}{3\,\bar{r}}\right)^{2/3}
-\bar{M}\left(\frac{l}{3\,\bar{r}}\right)^{4/3},\\
\bar{M}&=\left(\frac2{(\alpha^2+4)\left(\alpha+\sqrt{\alpha^2+4}\right)}\right)^{2/3}.
\end{align}
\end{subequations}
In this case, the Noether potential and the surface term yield
\begin{align*}
\int_{0}^{\bar{M}}\!\!\!\!d\bar{M}\Theta^{r}={}&
\frac{4\alpha\bar{M}^{1/2}}{{\kappa}l}
\left(\frac{3\,\bar{r}}{l}\right)^{2/3}
-\frac{(\alpha^2-28)\bar{M}}{{\kappa}l}\\
&-\frac{20\alpha\bar{M}^{3/2}}{{\kappa}l}
\left(\frac{l}{3\,\bar{r}}\right)^{2/3}
-\frac{16\bar{M}^{2}}{{\kappa}l}
\left(\frac{l}{3\,\bar{r}}\right)^{4/3},\nonumber\\
K^{rt}={}&-\frac{4\alpha\bar{M}^{1/2}}{{\kappa}l}
\left(\frac{3\,\bar{r}}{l}\right)^{2/3}
+\frac{4(\alpha^2-4)\bar{M}}{{\kappa}l}\\
&+\frac{20\alpha\bar{M}^{3/2}}{{\kappa}l}
\left(\frac{l}{3\,\bar{r}}\right)^{2/3}
+\frac{16\bar{M}^{2}}{{\kappa}l}
\left(\frac{l}{3\,\bar{r}}\right)^{4/3}.
\end{align*}
Finally, the rigid mass of the soliton is given by
\begin{equation}\label{massolsol3}
\mathcal{M}_{\mathrm{sol}}=\frac{12\pi(\alpha^2+4)^{1/3}}
{2^{1/3}\kappa\left(\alpha+\sqrt{\alpha^2+4}\right)^{2/3}},
\end{equation}
which again allows to rewrite the entropy (\ref{entropysoln3})
\emph{\`a la} Cardy, providing an additional realization of the
general formula (\ref{|Cardy|}).

\section{Conclusions}

In this paper, we confirm diverse general results concerning
Lifshitz black holes in new concrete examples. More precisely,
we are interested in identifying the mass of any reasonable
asymptotically Lifshitz configuration. For this, we
successfully tested the quasilocal formulation of conserved
charges recently proposed in \cite{Kim:2013zha,Gim:2014nba} and
based on the off-shell extension of the ADT formalism
\cite{Abbott:1981ff}. We focus our attention in the
three-dimensional case where the advantage lies in the fact
that our expressions for the masses can be checked, on one hand,
by using the first law of black hole thermodynamics and, on the
other hand, by independently verifying the anisotropic
generalization of the Cardy formula \cite{Gonzalez:2011nz}. In
order to achieve this task we supplement the action of new
massive gravity \cite{Bergshoeff:2009hq}, which already support
a Lifshitz black hole in vacuum \cite{AyonBeato:2009nh}, with
the one of a self-interacting scalar field nonminimally coupled
to gravity. Some of the authors have proven that this is a
useful strategy to enlarge the zoo of three-dimensional
Lifshitz black holes \cite{Correa:2014ika}. We start by
studying a family of solutions formerly found in
Ref.~\cite{Correa:2014ika} for generic dynamical exponent $z$
and characterized by a non-vanishing Wald entropy. Later, we
concentrate in the exponent $z=3$, relevant for the vacuum
\cite{AyonBeato:2009nh}, but excluded from the nontrivial
configurations exhibited in \cite{Correa:2014ika}. We found a
new family of Lifshitz black holes for a generic value of the
nonminimal coupling parameter $\xi$. Both families allows
massive and quartic contributions in their self-interactions,
however; the last solution is enhanced for $\xi=3/20$ by
turning on also a cubic contribution. We derived the black hole
mass of each of these solutions through the generalization of
the ADT formalism. The advantages of this method lie
essentially in the fact that the expression for the mass can be
obtained without assuming \emph{a priori} any asymptotic conditions,
without linearizing the equations of motion and uniquely
requiring to work out with the Lagrangian and the appropriate
Killing vector. We compute the Wald entropy and check that the
first law of black hole thermodynamics is valid in these three
cases for the obtained mass. In fact, all of them satisfy an
anisotropic version of the Smarr formula saying the mass is a
homogeneous function of the entropy with degree $z+1$. We
operate a completely independent verification of the results of
the method in three steps. We first derive the corresponding
soliton solution for the three different classes of solutions;
second we compute their respective mass using the same
quasilocal formulation of conserved charges. In order to apply
the method in these cases we use a one-parameter family of
solutions locally equivalent to the solitons that properly
have no integration constants, and consistently obtain a fixed
value for their masses. Finally, we confirm the validity of the
anisotropic generalization of the Cardy formula
(\ref{generCardy}), obtained from holographic arguments under
the assumption that the soliton plays in the bulk the role of
the ground state of a non-relativistic boundary theory
\cite{Gonzalez:2011nz}. The family (\ref{lifbhsoln1}) was
originally derived in Ref.~\cite{Correa:2014ika} together with
other two classes of Lifshitz black hole solutions. It is
simple to verify that the two remaining classes have a zero
Wald entropy. In the interest of performing a cross check of
the efficiency of the quasilocal method, we verify that
the generalized ADT formalism yields to a zero mass in these
cases, which again fits consistently with the first law and the
other tested formulas. In addition to the zero mass Lifshitz black holes produced in the studied theory, the three examples analyzed in the paper contain Lifshitz black holes with negative mass.
It is important to emphasize that our checking of the first law
is performed independently of the sign of the mass. The same
happens for the Smarr formula. Regarding the anisotropic Cardy
formula we point out a subtlety for these cases, where
additionally, the mass of the corresponding soliton is positive.
This invalidates the holographic interpretation provided in
Ref.~\cite{Correa:2014ika}; however, their semiclassical
arguments still apply which allow a general writing of their
formula involving the absolute values of the masses and a sign
correction in the entropy compatible with the first law. All
the cases under study are compatible with this general formula
(\ref{|Cardy|}). Returning to the Smarr formula, it would be
interesting to derive the expression (\ref{Smarr}) by
exploiting a scaling symmetry of the field equations, and to
derive the corresponding Noether conserved current. Evaluating
this latter at infinity and at the horizon and equating these
two expressions, the expectation is to obtain the anisotropic
Smarr formula, generalizing the results of
\cite{Banados:2005hm}. Finally, we would like to stress that it
is worth to pursue with testing this method by obtaining the
mass of higher-dimensional Lifshitz black holes. This task has
been started in \cite{Gim:2014nba} for special cases of the
vacuum configurations with square gravity corrections found in
\cite{AyonBeato:2010tm}, and for which it is known that the
naive extension of the ADT formalism does not work
\cite{Devecioglu:2010sf}. See, also, \cite{Fan:2014ala} for a
different perspective based on the first law.

\section{Acknowledgments}

This work has been partially funded by FONDECYT Grants No.
1121031, No. 1130423, No. 1141073, No. 11090281, No. 11121651, CONACYT Grants No. 175993,
178346, and CONICYT Grants No. 80130051, No. 79112034. This project was partially funded by Proyectos
CONICYT, Research Council UK (RCUK) Grant No. DPI20140053. E.A.-B.\ is supported
by ``Programa Atracci\'{o}n de Capital Humano Avanzado del
Extranjero, MEC'' from CONICYT. M.B-G.\ is supported by BECA
DOCTORAL CONICYT Grant No. 21120271. F.C.\ is also supported by the
Alexander von Humboldt Foundation. M.H.\ is also supported by
Departamento de Relaciones Internacionales ``Programa Regional
MATHAMSUD 13 MATH-05''. M.M.J-A.\ received financial support from
``Programa de Becas Nacionales'' of CONACYT, ``Plataforma de
Movilidad Estudiantil Alianza del Pac\'ifico'' of the Agencia
de Cooperaci\'on Internacional de Chile and ``Apoyo para
obtenci\'on de grado'' of CINVESTAV. CECs is funded by the
Chilean Government through the Centers of Excellence Base
Financing Program of CONICYT.

\end{document}